\documentstyle[12pt,epsfig]{article}
\begin{document}
\noindent
\begin{center}
{\Large {\bf Can Brans-Dicke Scalar Field Mimic \\Early Dark Energy?  }}\\
\vspace{2cm}
 ${\bf Yousef~Bisabr}$\footnote{e-mail:~y-bisabr@sru.ac.ir.}\\
\vspace{.5cm} {\small{Department of Physics, Shahid Rajaee Teacher
Training University,
Lavizan, Tehran 16788, Iran}}\\
\end{center}
\begin{abstract}
We show that a generalisation of Brand-Dicke (BD) theory can provide a theoretical basis for dark energy at early times. This generalisation is based on introducing a potential for the BD scalar and allowing of the latter to interact with radiation in the radiation-dominated rea.
We argue that the model we take up provides a strong theoretical basis for early dark energy (EDE) so that the scalar field can mimic a cosmological constant and quintessence behaviors under slow-roll and non-slow-roll conditions, respectively.

\end{abstract}
Keywords : Modified Gravity, Cosmology, Hubble Tension, Early Dark Energy.

~~~~~~~~~~~~~~~~~~~~~~~~~~~~~~~~~~~~~~~~~~~~~~~~~~~~~~~~~~~~~~~~~~~~~~~~~~~~~~~~~~~~~~~~~~~~~~~~~~~~~
\section{Introduction}
The standard cosmology is based on $\Lambda$CDM model which is a prevailing cosmological model describing the large-scale structure and dynamics of the universe.  It is based on two key components: 1) the cosmological constant ($\Lambda$)\cite{carroll} which represents the dark energy (DE) with equation of state parameter $\omega_{\Lambda}=-1$ and 2) the cold dark matter (CDM) which accounts for the dark matter (DM) component. The latter contains non-relativistic particles that do not interact with baryonic matter and radiation, making them invisible and detectable only through their gravitational effects.\\
Beside the successful role of $\Lambda$CDM in explaining the evolution of the Universe, there are also unexplained features in the standard cosmology. The fine tuning problem accounts for the sharp contradiction between theoretical predictions and observational estimates on $\Lambda$ which is of $120$ orders of magnitudes \cite{wein}. The coincidence problem concerns with the coincidence between the energy densities of DE and DM while they evolve differently as the Universe expands \cite{bis1}.\\
This puzzling picture has been recently enhanced by the so-called Hubble tension \cite{ht1}. It consists of
a discrepancy between $H_0$ obtained from local measurements and those inferred from the early Universe observations. This discrepancy ranges from $4.4\sigma$ to more than $6\sigma$ depending on the combination of the
local data used \cite{a}.
The local measurements involve measuring distances to nearby galaxies using Cepheid variable stars and Type Ia supernovae as standard candles. For instance, some local measurements are: 1) SH0ES Collaboration reported observations from the Hubble Space Telescope (HST) of Cepheid-calibration which gives $H_0 = 73.04 \pm 1.04~Kms^{-1} Mpc^{-1}$ \cite{local1}, 2) the value $H_0 = 72.2~\pm~2.1~Kms^{-1} Mpc^{-1}$ is reported by strong lens
systems and time-delay measurements \cite{local3}, 3) the Carnegie-Chicago Hubble program (CCHP) using the tip of the red giant branch (TRGB) method provides one of the most accurate and precise means of measuring the distances to nearby galaxies and gives the value $H_0 = 69.6~\pm~0.8 (\pm 1.1\%~ stat) \pm 1.7(\pm 2.4\% ~sys) Kms^{-1}Mpc^{-1}$ \cite{local2}.\\
 On the other hand, the early-time measurements are based on the Planck data of the CMB radiation yielding a lower value  $H_0 = 67.4 \pm 0.5 Kms^{-1}Mpc^{-1}$ \cite{agha}. Moreover, measurements based on baryon acoustic oscillations (BAO) and Large-scale Structure (LSS) gives $H_0
 =67.6\pm0.6~ Kms^{-1}Mpc^{-1}$ \cite{early1}, the arcminute-resolution maps of the CMB temperature and polarization anisotropy from the Atacama Cosmology Telescope (ACT) gives $H_0 =67.9\pm 1.5~ Kms^{-1}Mpc^{-1}$ \cite{early2}.\\
 There have been a number of proposals for solving the Hubble tension. The first one is based on the possibility that
Planck and/or the local distance ladder measurements
of $H_0$ suffer from unaccounted systematics \cite{sys}. However, no obvious systematic effect is found in
measurements and therefore there are now increasing attention on new physics beyond the standard $\Lambda$CDM model.
More intriguing possibility is existence of non-negligible DE contribution in the energy components of the Universe at
early times. Such early dark energy (EDE) could behave like a cosmological constant before some critical redshift but whose energy
density then dilutes faster than radiation \cite{ede}.
In addition to these proposals, modified gravity theories offer a promising avenue for resolving the tension
by changing the early-time dynamics of the Universe or redefining the distance ladder.  One may also consider a combination of the latter two proposals, namely an EDE generating by a modified gravitational theory. For instance, it is shown that according to recent data analyses of the supernovae Ia Pantheon sample, it is possible that $f(R)$ gravity theories can account for an effective redshift dependence of the Hubble constant \cite{fr}\cite{man}.\\
In this note we would like to use the same strategy and consider BD theory to formulate an EDE model. BD theory is endowed with a scalar degree of freedom, which is a non-minimally coupled scalar field interacting with matter sector in Einstein frame. We will use a generalised BD theory in which the BD scalar field dynamics is controlled by a general potential function. We then apply this theory to a radiation-dominated era at early Universe. Given the radiation-dominated era of interest, any contribution from dark matter (DM) are assumed to be negligible, allowing us to focus to the scalar field interaction with radiation as the primary source influencing cosmic evolution. The BD scalar field, through its coupling with radiation, introduces an effective fluid with distinct dynamical properties.\\
This work is organized as follows: In section 2, we introduce the model and derive the field equations and conservation laws in the Einstein frame. Due to the interaction of the BD scalar field with radiation, the corresponding stress-energy tensors are not separately conserved and there is an energy transfer between the two components.
In section 3, we apply this model to a spatially flat Friedman-Robertson Walker metric. In this case, we show that the (non-)conservation equation of radiation can be solved analytically.
We will show that if the energy transfer between radiation and the BD scalar field takes on a constant value, the BD scalar is then given in terms of logarithm of the scale factor so that the rate of change of the scalar field is closely related to the Hubble rate. In this case, we will show that the BD scalar field acts as EDE so that the equation of state parameter of the effective fluid takes those of a cosmological constant and quintessence in slow-roll and non-slow-roll dynamics, respectively.

~~~~~~~~~~~~~~~~~~~~~~~~~~~~~~~~~~~~~~~~~~~~~~~~~~~~~~~~~~~~~~~~~~~~~~~~~~~~~~~~~~~~~~~~~~~~~~~~~~~~~~~~~~~~~~~~~~~~~~~~~~~~~~~~
\section{The Model}
Let us begin with BD theory described by the action\footnote{We use units in which $8\pi G=M_p^{-2}=1$.}
\begin{equation}
S^{BD}_{JF}=\frac{1}{2}\int d^4x \sqrt{-\bar{g}} \{\phi
\bar{R}-\frac{\omega_{BD}}{\phi}\bar{g}^{\mu\nu}\bar{\nabla}_{\mu}\phi
\bar{\nabla}_{\nu}\phi -V(\phi)+2 L_{m}\}
\label{1}\end{equation} where $\bar{R}$ is the Ricci scalar, $\phi$
is a scalar field, $V(\phi)$ is a potential function and $L_{m}$ is Lagrangian density of matter sector which generally contains all possible matter or energy components in the Universe including baryonic matter, radiation and DM. The above action is the Jordan frame representation of the BD theory with the BD parameter $\omega_{BD}$. The potential $V(\phi)$ generalizes the BD theory which is shown to be relevant for studying some cosmological problems \cite{v}.  \\
A conformal transformation
\begin{equation}
\bar{g}_{\mu\nu}\rightarrow g_{\mu\nu}=\phi ~\bar{g}_{\mu\nu}
\label{a2}\end{equation} brings (\ref{1}) into the Einstein frame representation \cite{far1}.  One may redefine the scalar field by introducing
\begin{equation}
\varphi=\sqrt{\omega_{BD}+\frac{3}{2}}~\ln
\phi \label{a3}\end{equation}
This makes the action (\ref{1}) change to
\begin{equation}
S^{BD}_{EF}= \int d^{4}x \sqrt{-g} \{\frac{1}{2}R
-\frac{1}{2}g^{\mu\nu}\nabla_{\mu}\varphi
\nabla_{\nu}\varphi-U(\varphi)+e^{-\sigma\varphi}L_{m}\}
\label{a5}\end{equation} where
\begin{equation}
\sigma=8\sqrt{\frac{\pi}{2\omega_{BD}+3}}
\label{aaa5}\end{equation}
Here $\nabla_{\mu}$ is the covariant derivative of the
rescaled metric $g_{\mu\nu}$.  The Einstein frame potential is $U(\varphi)= V(\phi(\varphi))~e^{-\sigma\varphi}$. In the action (\ref{a5}) there is an interaction between $\varphi$ and the matter sector which has an essential role in our analysis. Moreover, the theory given by (\ref{a5}) is connected to more general modified gravity theories. For instance, (\ref{a5}) is equivalent to the Einstein frame representation of $f(R)$ Gravity models. In fact, metric $f(R)$ gravity
is dynamically equivalent to generalized BD theory  with a null BD parameter $\omega_{BD} = 0$ \cite{sot}.\\
Variation of the action (\ref{a5}) with
respect to the metric $g_{\mu\nu}$ and $\varphi$ gives,
respectively,
\begin{equation}
G_{\mu\nu}= T^{\varphi}_{\mu\nu}+e^{-\sigma\varphi}~ T^{m}_{\mu\nu}
\label{2}\end{equation}
\begin{equation}
\Box \varphi-\frac{dU(\varphi)}{d\varphi}=\sigma e^{-\sigma\varphi} L_{m} \label{3}\end{equation}
where
\begin{equation}
T^{\varphi}_{\mu\nu}=(\nabla_{\mu}\varphi
\nabla_{\nu}\varphi-\frac{1}{2} g_{\mu\nu}\nabla_{\alpha}\varphi
\nabla^{\alpha}\varphi) -U(\varphi)g_{\mu\nu}
\label{4}\end{equation}
\begin{equation}
T^{m}_{\mu\nu}=\frac{-2}{\sqrt{-g}}\frac{\delta (\sqrt{-g}L_{m})}{\delta
g^{\mu\nu}}
\label{5}\end{equation}
are the stress-tensors of $\varphi$ and the matter fields. Applying the Bianchi identities to (\ref{2}) indicates that $T_{\mu\nu}^{m}$ and $T_{\mu\nu}^{\varphi}$ do not separately conserved. It
gives
\begin{equation}
\nabla^{\mu}T_{\mu\nu}^{\varphi}=Q
 \label{3aa}\end{equation}
\begin{equation}
\nabla^{\mu}(e^{-\sigma\varphi}T_{\mu\nu}^{m})=-Q
 \label{3a}\end{equation}
where $Q$ is the interaction function. This function can be obtained by using (\ref{3}) and (\ref{4}) into (\ref{3aa}) which gives $Q= \sigma \nabla_{\nu}\varphi e^{-\sigma\varphi}L_{m}$.\\ We will
consider a perfect fluid description for the matter system with $\rho_{m}$ and $p_{m}$ being energy density and pressure, respectively, and $\omega_{m}\equiv p_{m}/\rho_{m}$ being the corresponding equation of state parameter.
Details of
the energy exchange between matter and $\varphi$ depends on the
explicit form of $L_{m}$. There are different choices for $L_m$
which all of them leads to the same energy-momentum tensor and field
equations in the context of general relativity \cite{haw} \cite{sh}.
Here we take $L_{m}=p_{m}$ for this
lagrangian density.

~~~~~~~~~~~~~~~~~~~~~~~~~~~~~~~~~~~~~~~~~~~~~~~~~~~~~~~~~~~~~~~~~~~~~~~~~~~~~~~~~~~~~~~~~~~~~~~~~~~~~~~~~~~~~~~~
\section{Cosmological implications}
We would like to apply the field equations (\ref{2}) and (\ref{3}) to the radiation-dominated era before recombination. In this case, the dominant component in $L_m$ is just radiation which is interacting with $\varphi$. Moreover, we use a spatially flat Friedmann-Robertson-Walker spacetime
 $
ds^2=-dt^2+a^2(t)(dx^2+dy^2+dz^2)$ with
$a(t)$ being the scale factor. In this case, the equations (\ref{2}) and (\ref{3}) become
\begin{equation}
3H^2=\rho_{eff}\equiv e^{-\sigma\varphi}\rho_r+\rho_{\varphi}
\label{a11}\end{equation}
\begin{equation}
2\dot{H}+3H^2=-p_{eff}\equiv -(e^{-\sigma\varphi}p_{r}+p_{\varphi})
\label{a12}\end{equation}
\begin{equation}
\ddot{\varphi}+3H\dot{\varphi}+\frac{dU(\varphi)}{d\varphi}=-\sigma e^{-\sigma\varphi}p_{r}
\label{a13}\end{equation}
where $H=\frac{\dot{a}}{a}$ is the Hubble parameter. The conservation equations
become
\begin{equation}
\dot{\rho}_{\varphi}+3H(\omega_{\varphi}+1)\rho_{\varphi}=-\sigma e^{-\sigma\varphi}\dot{\varphi}p_{r}
\label{a15}\end{equation}
\begin{equation}
\dot{\rho}_{r}+4H\rho_{4}=\frac{4}{3}\sigma  \dot{\varphi}
\rho_{r} \label{a14}\end{equation}
 The latter can be immediately solved which gives the following solution
\begin{equation}
\rho_r= \rho_{0r}a^{-4}e^{\frac{4}{3}\sigma\varphi}\label{a16}\end{equation} where
$\rho_{0r}$ is an integration constant. This solution can also be written as
 \cite{bis1}
\begin{equation}
\rho_{r}=\rho_{0r}a^{-4+\varepsilon}
\label{a166}\end{equation} where we have defined
\begin{equation}
\varepsilon\equiv \frac{4\lambda}{3}\frac{\sigma\varphi}{\ln a}\label{a16}\end{equation}
with $\lambda$ being a free parameter. This solution indicates that the
evolution of energy density is modified due to interaction of
$\varphi$ with radiation.  For $\varepsilon>0$, radiation
is created. In this case, energy is injecting from $\varphi$ into radiation so that the latter dilutes more slowly compared to the standard evolution
$\rho_{r}=\rho_{0r} a^{-4}$. For $\varepsilon<0$, on the other hand, radiation is annihilated and energy transfers outside of radiation so that the rate of dilution of $\rho_{r}$ is faster than the standard one.\\
The equations (\ref{a15}) and (\ref{a14}) indicate energy exchange between $\varphi$ and radiation which modifies evolution of $\rho_{r}$ and $\rho_{\varphi}$. In this interacting system, $\varepsilon$ in (\ref{a166}) is generally an evolving function characterizing the rate of the energy transfer. In the case that $\varepsilon$ can
be regarded as a constant parameter (\ref{a16}) reduces to
\begin{equation}
\varphi=\gamma \ln a
\label{c2}\end{equation}
with $\gamma\equiv\frac{3\varepsilon}{4\lambda\sigma}$. Even though the constancy of $\varepsilon$ is not generally true, it may  however hold during periods of the expansion history and simplifies greatly our analysis.
The relation (\ref{c2}) implies that the rate of change of the scalar field is given by the Hubble rate $\dot{\varphi}=\gamma H$\footnote{Theoretically, one may move away from the logarithmic profile for $\phi(a)$, however, this would introduce more complex potentials, likely making it challenging to maintain the straightforward connection between $\dot{\phi}$ and $H$. }.
Combining this relation with (\ref{a11}) and (\ref{a13}) leads to the consistency relation
\begin{equation}
\dot{H}+3 H^2+ \frac{1}{\gamma}\frac{dU}{d\varphi}=-\frac{\sigma}{\gamma}p_{r} (\frac{a}{a_c})^{-\sigma\gamma}
\end{equation}
which must be satisfied to ensure that (\ref{c2}) is consistent with the solutions. It is shown that such a consistency holds for exponential potentials \cite{exp}.\\
The equation of state parameter of the effective fluid evolves as
\begin{equation}
\omega_{eff}\equiv\frac{p_{eff}}{\rho_{eff}}=\frac{e^{-\sigma\varphi}p_{r}+p_{\varphi}}{e^{-\sigma\varphi}\rho_{r}+\rho_{\varphi}}
\label{om}\end{equation}
An interesting feature of (\ref{om}) is the exponential coefficient $e^{-\sigma\varphi}$ in front of $p_{r}$ and $\rho_{r}$. In the non-interacting limit ($\varepsilon=0$), $\rho_{r}\propto a^{-4}$ goes to infinity as $a\rightarrow 0$. On the other hand, in the interacting case ($\varepsilon\neq 0$) $\varphi$ is a time-dependent field which goes to infinity when $a \rightarrow 0$ as it is evident from (\ref{c2}). In this case, the coefficient $e^{-\sigma\varphi}$ acts as a damping factor which suppresses the growth of $\rho_{r}$ and $p_{r}$ at early times. The behaviors  of $\varphi$, $\rho_{r}$ and $e^{-\sigma\varphi}\rho_{r}$ are plotted in figs. 1, 2 which shows that the latter diminishes  as $a \rightarrow 0$. \\The vanishing of $e^{-\sigma\varphi}\rho_{r}$ at early times gives an interesting feature to $\omega_{eff}$. If $\varphi$ is a slow-roll field with a potential satisfying  $\frac{1}{2}\dot{\varphi}^2<<V(\varphi)$, (\ref{om}) reduces to
\begin{equation}
\omega_{eff}=\frac{e^{-\sigma\varphi}p_{r}-U(\varphi)}{e^{-\sigma\varphi}\rho_{r}+U(\varphi)}
\label{om2}\end{equation}
which gives $\omega_{eff}\rightarrow -1$ as $a\rightarrow 0$. If $\varphi$ is a non-slow-roll field, on the other hand, we may use (\ref{a11}) and (\ref{c2}) to obtain
\begin{equation}
\omega_{eff}=\frac{(\frac{1}{3}+\delta_1) e^{-\sigma\varphi}\rho_{r}+\delta_2U(\varphi)}{\delta_1(e^{-\sigma\varphi}\rho_{r}+U(\varphi))}
\end{equation}
where
\begin{equation}
\delta_1=\frac{\frac{1}{2}\gamma^2}{(3-\frac{1}{2}\gamma^2)}
\end{equation}
\begin{equation}
\delta_2=\frac{\gamma^2-3}{(3-\frac{1}{2}\gamma^2)}
\end{equation}
Again $e^{-\sigma\varphi}\rho_{r}\rightarrow 0$ as $a\rightarrow 0$, and $\omega_{eff}$ takes a constant value $\omega_{eff}\rightarrow \frac{\delta_2}{\delta_1}=\frac{1}{3}\gamma^2-1$ which is bounded by $-1<\omega_{eff}<0$ for $\gamma<1$. In this case, $\varphi$ appears as a quintessence field at early times.
~~~~~~~~~~~~~~~~~~~~~~~~~~~~~~~~~~~~~~~~~~~~~~~~~~~~~~~~~~~~~~~~~~~~~~~~~~~~~~~~~~~~~~~~~~~~~~~~~~~~~~~~~~~~~~
\section{Discussion}
The interaction between the BD scalar field and radiation during the early Universe, particularly in the radiation-dominated era, leads to several changes in CMB anisotropies and the overall structure formation as  briefly explained in the following:\\
\textbf{1) altered expansion rate and sound horizon}\\
The BD scalar field's interaction with radiation changes the effective energy density in the early Universe. This modification impacts the expansion rate, resulting in a shift in the Hubble parameter during the radiation-dominated era. Consequently, the change in the expansion rate affects the sound horizon scale at the time of recombination, which is crucial for determining the spacing of acoustic peaks in the CMB power spectrum.
If the BD scalar field behaves like a cosmological constant during this period, particularly under slow-roll conditions, it could have a freezing effect on the fluctuations \cite{free}. This influence may be reflected in shifts in the peak locations of the CMB temperature power spectrum, especially in the first few acoustic peaks.\\
\textbf{2) modifications to acoustic oscillations}\\
The $\varphi$-radiation interaction also alters the oscillatory behavior of the photon-baryon plasma. This interaction serves as an additional source of energy density, leading to slight delays in the onset of matter-radiation equality and affecting the amplitude and height of the acoustic peaks. Specifically, the energy density from the BD scalar could either enhance or suppress the first acoustic peak, depending on the details of the scalar-radiation coupling. Moreover, the oscillatory behavior of the CMB peaks is also influenced by this interaction. In cases where the BD field evolves slowly, this effective freezing could increase.\\
\textbf{3) integrated sachs-wolfe (ISW) effect}\\
The coupling of the BD scalar field could create variations in the gravitational potential, leaving a detectable ISW imprint. This would be evident in CMB correlations with large-scale structure, where changes in the rate of structure growth lead to ISW anomalies, particularly in the low-$\ell$ regions of the CMB power spectrum.\\
\textbf{4) enhanced structure growth and early matter-radiation equality}\\
The additional density component from the scalar field interaction may influence the growth rate of small-scale perturbations. This could be observed in large-scale structure data, as EDE tends to slow down structure growth, resulting in fewer small structures than what models without such interactions would predict.
~~~~~~~~~~~~~~~~~~~~~~~~~~~~~~~~~~~~~~~~~~~~~~~~~~~~~~~~~~~~~~~~~~~~~~~~~~~~~~~~~~~~~~~~~~~~~~~~~~~~~~~~~~~~~
\section{Conclusions}
In this study, we have explored BD theory with a potential in the Einstein frame as a theoretical model for EDE in the radiation-dominated era. This model is based on the interaction between the BD scalar field $\varphi$ and the matter Lagrangian density $L_{m}$ which was dominated by radiation in the radiation-dominated era. By analyzing the dynamics of this interaction, we have demonstrated how the behavior of $\varphi$ influences the effective equation of state parameter $\omega_{eff}$.
In particular, the interaction alters the behavior of the effective energy density and pressure in the radiation-dominated era. Specifically, as the scale factor approaches zero, the effective energy density and pressure asymptotically approach finite values, effectively vanishing in the limit. This behavior contrasts sharply with scenarios lacking such interaction, where the corresponding quantities diverge to infinity as the scale factor diminishes. These results highlight the profound impact of the interaction between the BD scalar and radiation, offering novel insights into the dynamics of EDE.\\
Our findings indicate that when slow-roll conditions hold, the effective equation of state parameter reaches a value of $\omega_{eff}=-1$. This result is significant as it suggests that at early times, such a slow-rolling scalar field can induce a phase of accelerated expansion, consistent with EDE models. Conversely, when the slow-roll conditions do not hold, $\omega_{eff}$ is bounded by $-1<\omega_{eff}<0$ indicating a quintessence behavior of the BD scalar field. These results contribute to a deeper understanding of the role of BD scalar in the evolution of the early Universe and offer a viable theoretical framework for explaining the dynamics of EDE and the Hubble tension.

~~~~~~~~~~~~~~~~~~~~~~~~~~~~~~~~~~~~~~~~~~~~~~~~~~~~~~~~~~~~~~~~~~~~~~~~~~~~~~~~~~~~~~~~~~~~~~~~~~~~~~~~~~~~~~~~

\begin{figure}[ht]
\begin{center}
\includegraphics[width=0.6\linewidth]{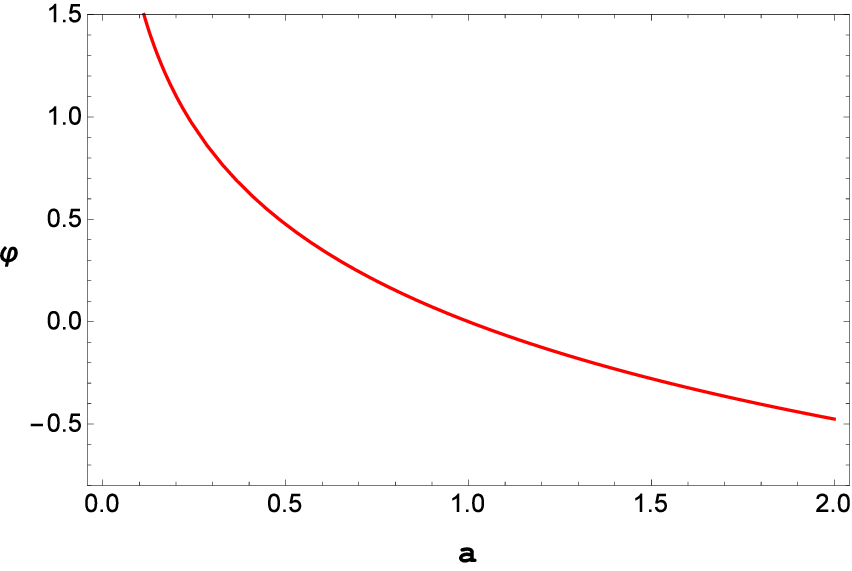}
\caption{ The plot of $\varphi$ against the scale factor $a$ for $\varepsilon=-0.6$ and $\lambda=0.08$.}
\end{center}
\end{figure}
\begin{figure}[ht]
\begin{center}
\includegraphics[width=0.6\linewidth]{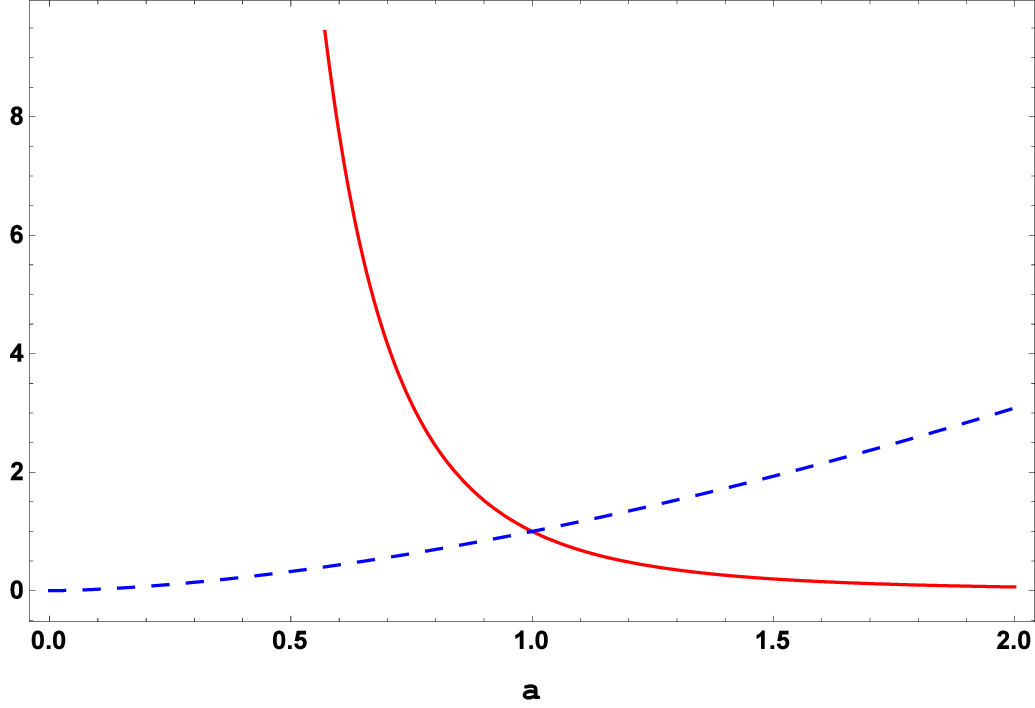}
\caption{ The plot of $\rho_r$ (the solid line) and $e^{-\sigma\varphi}\rho_r$ (the dashed line) against the scale factor $a$ for $\varepsilon=-0.6$ and $\lambda=0.08$. }
\end{center}
\end{figure}

\end{document}